# Multi-Functional Variable Thickness Structure for Broadband and Omnidirectional Focusing and Collimation

**Liuxian Zhao[a], Changquan Lai[b], Miao Yu[a,c,*]**

[a] Institute for Systems Research, University of Maryland, College Park, MD, 20742, USA

[b] Temasek Laboratories, Nanyang Technological University, 50 Nanyang Drive, Singapore 637553

[c] Department of Mechanical Engineering, University of Maryland, College Park, Maryland 20742, USA

*Author to whom correspondence should be addressed: mmyu@umd.edu

**ABSTRACT**

Luneburg lens is a symmetric gradient-index lens with a refractive index that increases from the outer surface to the center in a radial manner. It has the ability to focus and collimate waves, which makes it useful for energy harvesting, waveguiding and as a component in transducers. An ideal Luneburg lens should be easy to fabricate, has broadband and omnidirectional characteristics, as well as a focal length that can be easily tuned. However, existing structural Luneburg lenses based on phononic crystals can hardly achieve these requirements. Here, we propose an alternative structural Luneburg lens which has a refractive index that varies smoothly with its radial distance as a result of a changing thickness. Theoretical calculations, numerical simulations and experimental measurements of flexural wave propagation through the lens showed that flexural wave focusing can be obtained inside, at the edge and outside of the variable thickness lens for different frequencies and propagation directions. Flexural wave collimation was also demonstrated when a point source was placed at the respective focal points for each lens. Furthermore, it was shown that flexural waves that were focused onto a piezoelectric energy harvester by the Luneburg lens can lead to a significant increase in the harvested voltage compared to that obtained without focusing.

# 1. Introduction

There is widespread and sustained interest in the control and manipulation of structural waves for applications such as vibration-based energy harvesting [1], structural health monitoring [2], and medical diagnosis and therapy [3]. For example, focusing of structural waves can be an effective approach for enhancing vibration-based energy harvesting efficiency using piezoelectric transducers [4]. Furthermore, strong collimation of structural waves that can enable high performance acoustic transducers is highly desirable for ultrasonic imaging, diagnosis, and treatment [5, 6].

With the development of gradient-index (GRIN) acoustic metamaterials, focusing and collimation of elastic waves in thin plates have been achieved by using structural Luneburg lens [7-9]. For example, a simple GRIN structural Luneburg lens can be created by using an array of holes with varying radii to achieve the desired gradient refractive index [10]. Although the frequency response of these GRIN Luneburg Lenses was shown to be generally broadband, the performance of these devices was limited by the discrete nature of the hole lattice, which resulted in a stepwise refractive index profile. For the change in refractive index to appear smooth, the holes must be made much smaller than the wavelength, but this quickly becomes impractical as the frequency of the manipulated wave increases [11].

It is well-known that the local-dependence of the modulus can be achieved by varying the structural thickness to tailor the dispersion of flexural wave propagation in the structure. This approach, which is referred to as Acoustic Black Holes (ABH), has been employed for vibration control [12-15] and energy harvesting [16-18]. Since the refractive index depends not only on the material properties, but also on the structural thickness, smooth variations in refractive indices can be obtained with ABH. This approach has been extensively investigated for achieving efficient damping of flexural waves [19-21]. However, fabrication of an ideal

ABH is not possible due to the zero-thickness requirement (singularity point) at its center. To circumvent this limitation, Climente *et. al*. [22] proposed a structural Luneburg lens based on a thickness profile that is different from that of the ABH, and demonstrated its flexural wave focusing property through numerical simulations. However, the lens with such a thickness profile could only be used to focus the flexural waves at its edge.

In this study, we extend Climente *et. al.*'s initial work and developed a general formulation, based on studies of modified optical Luneburg lenses [23, 24], that relates the thickness profile of a structural Luneburg lens to its focal length, so that the focal point can be tailored to be anywhere inside, outside, or on the edge of the lens. The flexural wave focusing and collimation properties of these structural Luneburg lenses are then investigated using analytical, numerical, and experimental means, and its usefulness in energy harvesting will also be numerically and experimentally demonstrated.

## 2. Modified Structural Luneburg Lens Design

The modified structural Luneburg lens is based on a variable thickness structure defined in a thin plate (see inset of Figure 1(a)). As shown in Figure 1 (a), when a line source is used to excite structural waves in the plate, the flexural waves passing through the Luneburg lens will generate a focus. Conversely, when a point source is placed at the focal point for excitation, a collimated beam can be generated on the opposite side of the Luneburg lens. Figure 1 (b) shows the schematic of the variable thickness structure proposed in this work. The structure consists of a circular variable thickness region of radius $R$ defined in a uniform plate with a thickness of $h_0$.

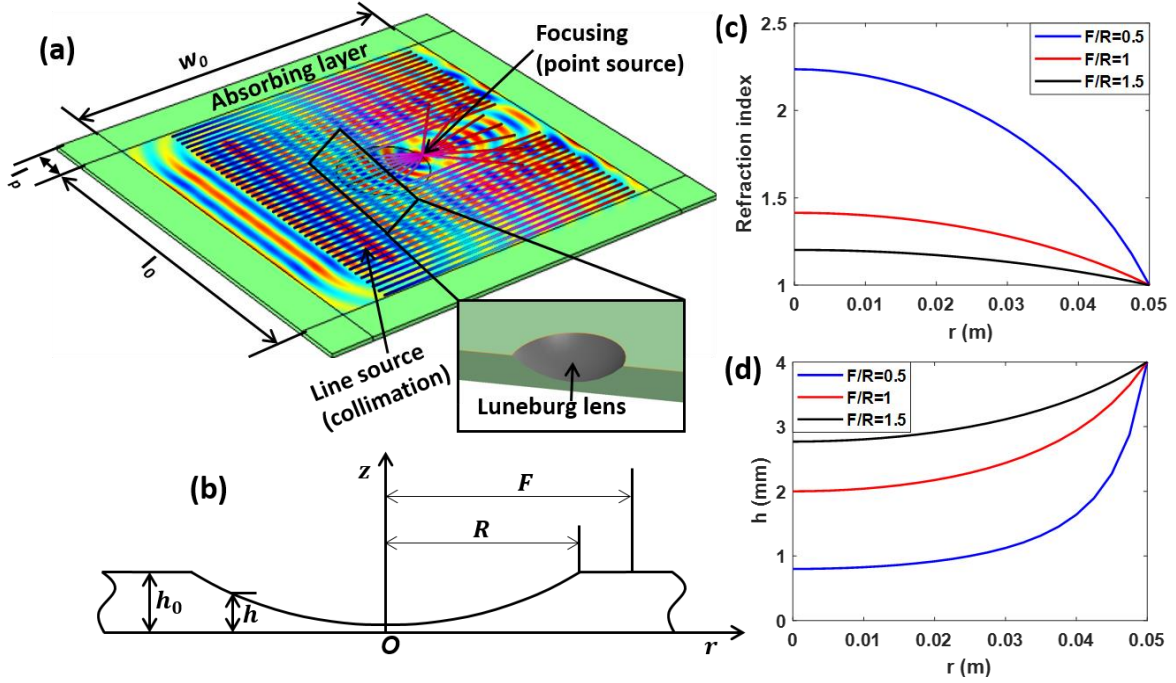

**Figure 1: Mechanism of the modified structural Luneburg lens and its design principles. (a) Schematic of the modified structural Luneburg lens for manipulating flexural waves (focusing and collimation). The inset shows a cut-away of the Luneburg lens, revealing its variable depth. (b) Cross-sectional view of the lens based on a variable thickness structure. $h(r)$ refers to the thickness profile at a radial distance, $r$, from the lens center, $O$, and $F$ is the focal length. (c) Refractive index and (d) thickness, $h$, as a function of radial distance, $r$, for different focal length ratios, $F/R$.**

Based on the modified optical Luneburg lens design [24], for the modified structural Luneburg lens shown in Figure 1, the graded refractive index, $n(r)$, should satisfy the following equation:

$$n(r) = \frac{\sqrt{F^2+R^2-r^2}}{F}, \qquad (1)$$

where $r$ is the radial distance from the lens center and $F$ is the focusing length. This equation essentially informs how the refractive index profile can be tailored with respect to the desired focal length, which may be anywhere inside, outside, or on the edge of the lens).

As a flexural wave propagates through the modified structural Luneburg lens, its phase velocity, c, is given as [16]:

$$c = \frac{\omega}{k} = \left(\frac{Eh^2\omega^2}{12\rho(1-v^2)}\right)^{\frac{1}{4}} \quad , \tag{2}$$

where $\omega$ is the angular frequency, $k$ is the wavenumber, $E$ is the Young's modulus, $\rho$ is the density and $v$ is the Poisson ratio. Since $n = \frac{c_0}{c}$, where $c_0$ refers to the phase velocity of the flexural wave passing through the flat plate of thickness $h_0$, the refractive index can be obtained from Equation (2) as

$$n = \sqrt{\frac{h_0}{h}} \quad . \tag{3}$$

Substituting Equation (3) into (1), the thickness profile can then be derived as:

$$h(r) = \frac{h_0 F^2}{F^2 + R^2 - r^2} \quad . \tag{4}$$

Equation (4) therefore allows one to derive the thickness profile for focusing flexural waves onto desired location. As a proof-of-concept, we choose the following parameters in our analytical, numerical, and experimental studies: $R = 0.05$ m and $h_0 = 0.004$ m. Three different cases, with focal length ratios, *F/R*, of 0.5 (focal point inside lens), 1 (focal point on edge of lens) and 1.5 (focal point outside lens) were investigated. The refractive indices and thickness profiles of these modified Luneburg lenses are shown in Figure 1 (c) and (d) respectively.

## 3. Focusing and Collimation

### 3.1 Analytical model

Based on the Snell's law, the refraction angle of a flexural ray is determined by the refractive index ratio between the two media. For flexural wave propagation over a two-dimensional lens, a geometrical acoustics approach in Hamiltonian formulation is used [13]. This approach allows a simple analysis of flexural wave propagation in plates with a cylindrically symmetrical profile, in which there is no geometrically induced anisotropy of flexural wave velocity. The governing equations defining the ray trajectory of a flexural wave inside a cylindrically symmetric tapered profile can be written in the form [25]:

$$\begin{cases} \frac{dr}{d\theta} = r \frac{1}{\tan(\alpha)} \\ \frac{d\alpha}{d\theta} = -1 - \frac{r}{n}\frac{dn}{dr} \\ nr\sin(\alpha) = \text{const} \end{cases}, \qquad (5)$$

where $r$ and $\theta$ are the polar coordinates with the origin at the centre of the structure ($r = \sqrt{x^2 + y^2}$, and $\theta = \arctan\left(\frac{y}{x}\right)$), $\alpha$ is the angle between $\theta = 0$ and the wave vector, $k$. Figure 2(a) shows the ray trajectories of flexural waves with $\alpha_1$ and $\alpha_2$ propagating through a lens (grey area). The ray with the incident angle $\alpha_1$ is bent when it interacts with the Luneburg lens and propagates forward, while the ray with the incident angle $\alpha_2$ continues as a straight line without interacting with the Luneburg lens.

The ray tracing method is used to study flexural wave focusing with the modified structural Luneburg lens. For focusing, a bundle of incident rays with frequency $f = 40$ kHz from a plane wave excitation propagate in the constant plate section along the $x$ axis from $-x$ towards $+x$ and interacts with modified structural Luneburg lens. The excitation plane wave is located at $x = -2R$, with a width of $3R$ (from $y = -1.5R$ to $y = 1.5R$). The calculated ray trajectories are shown in Figure 2 (b), (c) and (d) for $F/R = 0.5$, $F/R = 1$, and $F/R = 1.5$ respectively. Each of these $F/R$ ratios corresponds to a specific focus location at $(0.5R, 0)$, $(R, 0)$, and $(1.5R, 0)$. Note that there are some aberrations for $F > R$. These ray diagrams also indicate that a point source placed at the focal point for excitation can lead to collimated waves at the opposite side of the lens.

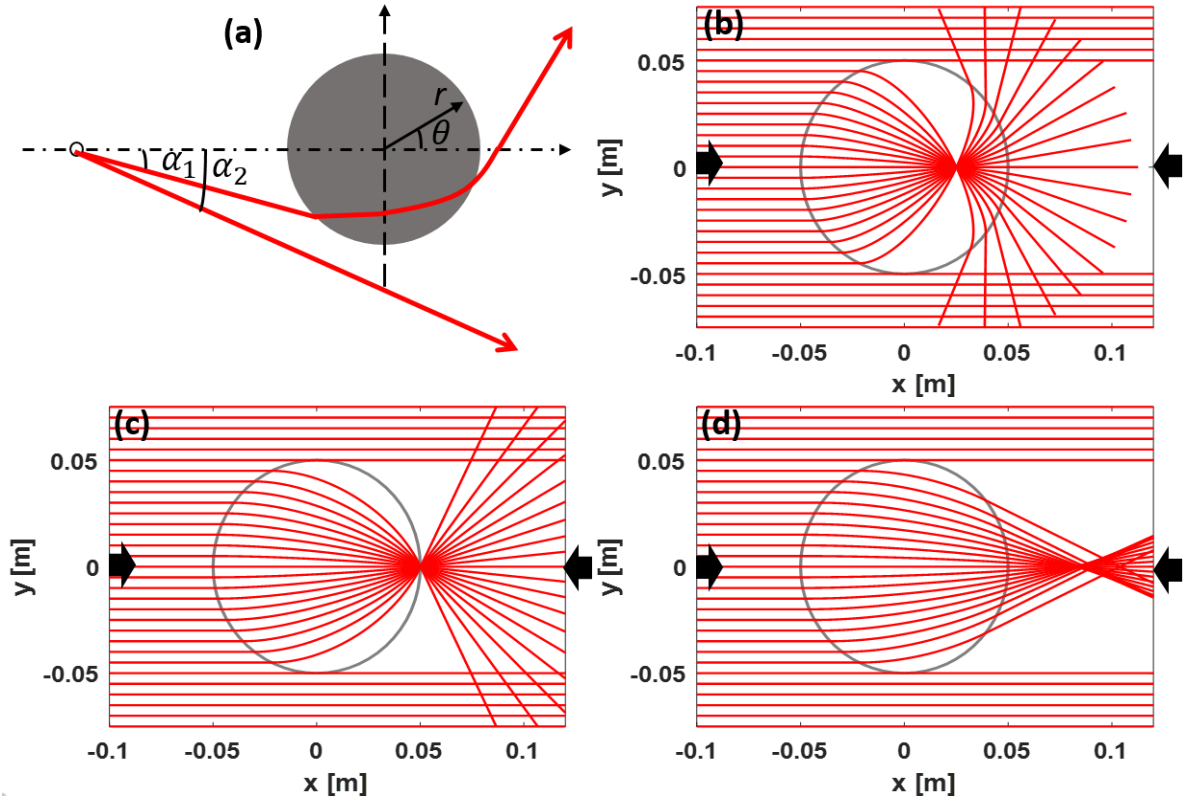

**Figure 2: Theoretical calculation of ray trajectories based on geometrical acoustic theory. (a)** Schematic of two typical ray trajectories, illustrating wave propagation over a modified structural Luneburg lens. Calculated ray trajectories for focal length ratios of **(b)** $F/R = 0.5$, **(c)** $F/R = 1$, and **(d)** $F/R = 1.5$. The right direction arrow indicates a plane wave excitation direction, and the left direction arrow indicates a point excitation direction. The grey circle indicates the outline of the Luneburg lens.

In order to demonstrate the omnidirectional characteristic, the ray trajectories are also calculated for a family of incident rays with frequency $f = 40$ kHz from a plane wave excitation at an incident angle of 45º. Figure S1 (a)-(c) show the ray trajectories for flexural wave focusing and collimation for $F/R = 0.5$, $F/R = 1$, and $F/R = 1.5$ respectively. These results demonstrate that the focusing and collimation properties of the modified structural Luneburg lens are omnidirectional.

### 3.2 Numerical simulations

Next, we perform numerical simulations to investigate flexural wave focusing and collimation by using the modified structural Luneburg lens (Figure 1) defined in a thin, rectangular aluminium plate. The material properties of the plate were as follows: density $\rho = 2700$ kg/m$^3$, Young's modulus $E = 70$ GPa, and Poisson's ratio $v = 0.33$. The dimension of the plate was $l_0 \times w_0 \times h_0 = 0.317$ m $\times$ 0.317 m $\times$ 0.004 m. Both frequency and time domain analyses were carried out. In the frequency domain analysis, Perfectly Matched Layers (PML) with a width $l_p = 0.07$ m were used to cover the four sides of the plate in order to reduce the boundary reflections (see Figure 1). In the time domain analysis, Low Reflecting Boundaries (LRB) were used to reduce the boundary reflections, as shown in Figure S2.

In order to verify the flexural wave focusing property of the lens at different locations, numerical simulations were performed for different focal length ratios of $F/R = 0.5$, $F/R = 1$ and $F/R = 1.5$. A line source ($x = -0.08$ m, and $y = -R$ to $y = R$) with an incident angle of 0 degree was used for excitation at a frequency of $f = 40$ kHz. The full field wave propagations are shown in Figure 3 (a)-(c), which clearly demonstrate that flexural wave focusing inside, on the edge, and outside of the modified structural Luneburg lens respectively. For flexural wave collimation, a point source excitation with an incident angle of 0 degree at the frequency of $f = 40$ kHz was generated at three locations corresponding to $F/R = 0.5$, $F/R = 1$, and $F/R = 1.5$. The simulation results in Figure 3 (d)-(f) clearly show the collimation property of the lens. The broadband and omnidirectional characteristics of modified structural Luneburg lens were also verified and the results are shown in Figure S3 and Figure S4, respectively.

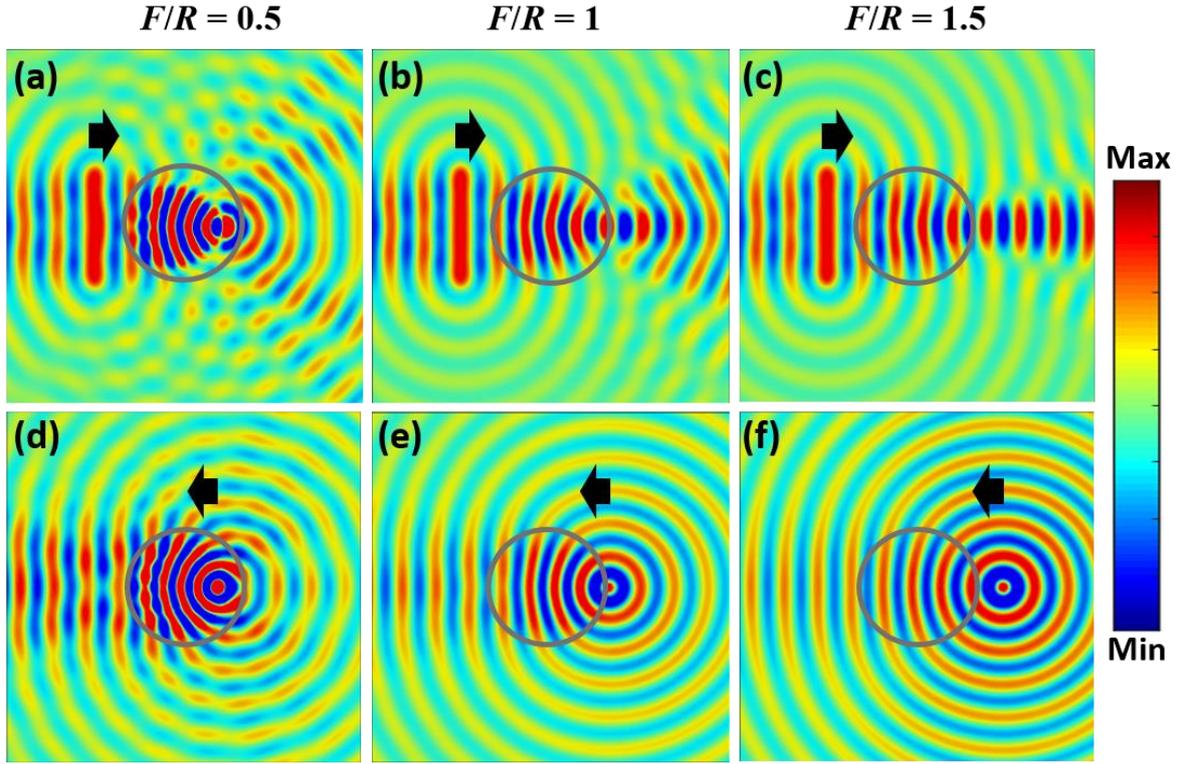

**Figure 3: Numerical simulations of the steady state response of the modified Luneburg Lens for flexural wave focusing and collimation. Flexural wave focusing with focal length ratios of (a) *F/R* = 0.5 (b) *F/R* = 1 and (c) *F/R* = 1.5, respectively. Flexural wave collimation with focal length ratios of (a) *F/R* = 0.5 (b) *F/R* = 1 and (c) *F/R* = 1.5, respectively. The grey circle indicates the outline of the Luneburg lens, and the color bar indicates the displacement. The right direction arrow indicates a plane wave excitation direction, and the left direction arrow indicates a point excitation direction.**

In order to show the flexural wave focusing at different time steps, time domain analysis was performed for the case of *F/R* = 1. A line source with an incident angle of 0 degree was used for excitation. The input to the source was a signal of 3-count tone bursts at *f* = 40 kHz. The obtained waveforms at different time instants of *t* = 0.03, 0.065, 0.085, and 0.105 ms are shown in Figure 4 (a)-(d). From 0 ms to 0.03 ms, the generated flexural waves propagated forward. After 0.03 ms, the flexural waves interacted with the Luneburg lens and the wavefronts gradually became narrower. From 0.065 to 0.085 ms, the flexural waves were localized at the edge of the lens to form the focus. After 0.085 ms, the focused waves propagated forward with a cylindrical wavefront and the amplitude of focusing waves remained at a high level.

Flexural wave collimation at different time steps was also numerically investigated. A point source was used for excitation at the focal point, which had the same properties as the line source described previously. The full field wave propagations at different time instants of 0.04, 0.075, 0.1, and 0.12 ms are shown in Figure 4 (e)-(h). From 0 to 0.04 ms, the generated flexural waves propagated from the focal point with a cylindrical wavefront. After 0.04 ms, flexural waves interacted with the modified structural Luneburg lens and the wavefront became flat. From 0.075 to 0.10 ms, the flexural waves propagated as a plane wave. After 0.1 ms, the plane wave exited the lens and continued to propagate with a plane wavefront. The omnidirectional characteristic of flexural wave focusing and collimation in the time domain was also verified and the results are provided in Figure S5.

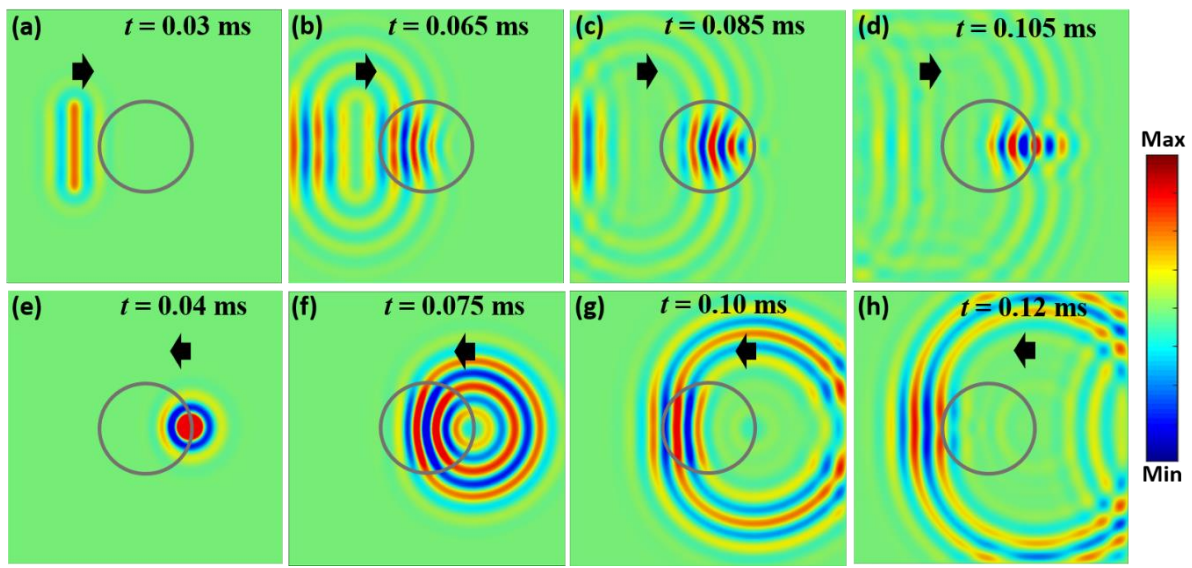

**Figure 4: Numerical simulations of transient response of the modified Luneburg Lens for flexural wave focusing and collimation with a focal length ratio $F/R = 1$. (a) - (d) are the flexural wave focusing at time instants of $t = 0.03$ ms, $t = 0.065$ ms, $t = 0.085$ ms, and $t =$**

**0.105 ms. (e) - (h) are the flexural wave collimation at time instants of $t = 0.04$ ms, $t = 0.075$ ms, $t = 0.10$ ms, and (d) $t = 0.12$ ms. The grey circle indicates the outline of the Luneburg lens, and the color bar indicates the displacement. The right direction arrow indicates a plane wave excitation direction, and the left direction arrow indicates a point excitation direction.**

### 3.3 Experimental studies

Experimental studies were also carried out to verify the focusing and collimation properties of the modified structural Luneburg lens. The experimental setup and the fabricated structural Luneburg lens in a plate are shown in Figure 5. The lens was fabricated on a thin 6061 aluminium plate (McMaster-Carr) with dimensions of 0.457 m × 0.457 m × 0.004 m. The four sides of the plate were covered with damping foil tapes (McMaster-Carr) with a width of $l_p = 0.07$ m. Therefore, the effective dimensions of the plate were 0.317 m × 0.317 m × 0.004 m, and the same as that of the numerical models. The radius of the structural Luneburg lens was $R = 0.05$ m.

Five rectangular piezoelectric transducers (dimensions 20 mm × 15 mm × 1 mm from STEMiNC Corp.) were used to generate a line source excitation and a circular piezoelectric disc (12 mm in diameter and 0.6mm in thickness from STEMiNC Corp.) was used to generate a point source excitation. The piezoelectric transducers were bonded to the plate by using 2P-10 adhesive (from Fastcap, LLC). For the experimental study, two types of modified structural Luneburg lens with focal length ratios of $F/R = 1$ and $F/R = 1.5$ were fabricated. Another identical plate instrumented with piezoelectric transducers but without the Luneburg lens was measured as a reference.

During the experiments, the two vertical sides of the plate were fixed on a frame, as shown in Figure 5 inset (a). A close-up view of the Luneburg lens is shown in Figure 5 inset (b). A scanning laser Doppler vibrometer (SLDV) (Polytec PSV-400) was used to measure the propagating wave field by recording the out-of-plane component of the particle velocity on the plate over the scan area shown in Figure 5. The scanning was performed on the flat side of the plate. Both frequency and time domain analyses were performed to characterize the flexural wave focusing and collimation properties of the modified structural Luneburg lens. For the frequency domain measurements, a continuous sinusoidal burst at frequency $f$ = 40 kHz was used for excitation, and for the time domain, the excitation source was a signal of 3-count tone bursts at the frequency, $f$ = 40 kHz.

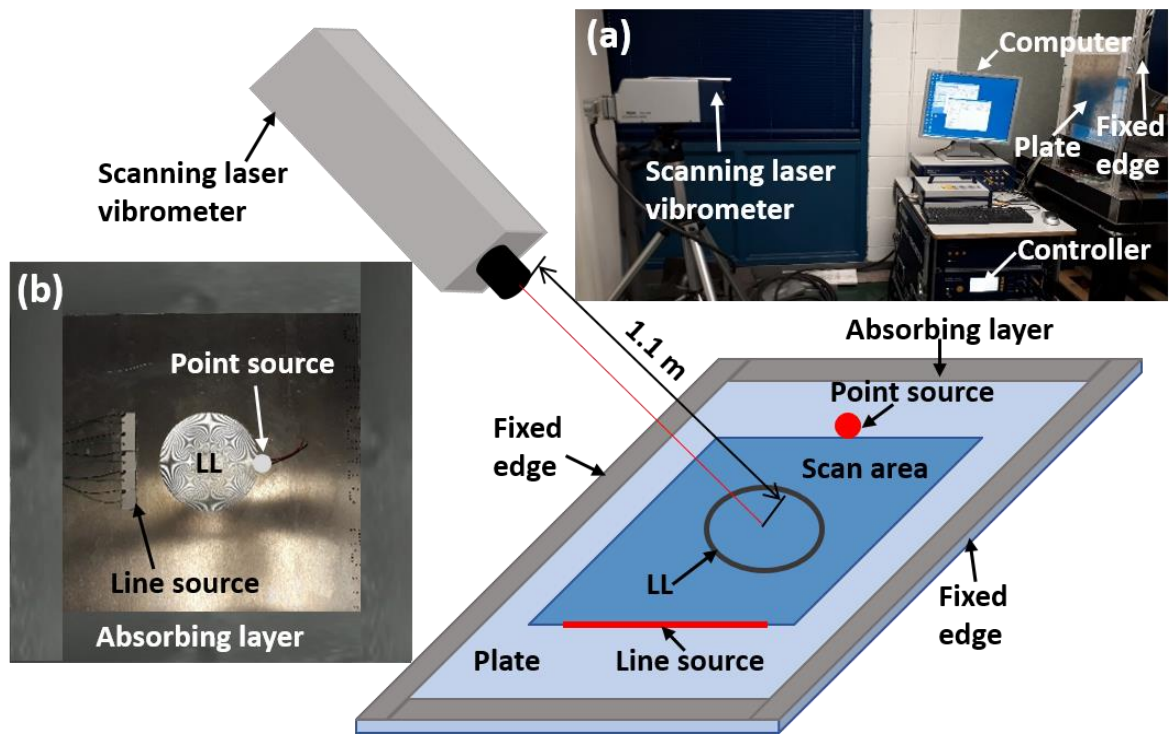

**Figure 5: Schematic of the experimental setup. A scanning laser vibrometer was used to measure the full-field wave propagations in the scan area. Inset (a) shows a photo of the experimental setup. The thin plate was constrained by two vertical fixtures and covered with absorbing layers. Inset (b) illustrates the fabricated Luneburg lens (LL) in the plate.**

The measured waveforms of the out-of-plane displacement fields on the reference plate with the line source and point source excitations are shown in Figures 6 (a) and (b), respectively. The results clearly show that the line source generates a plane wave, while the point source generates a circular wave on the plate. The measured waveforms of the out of plane displacement fields on the plate with the Luneburg lens ($F/R = 1$) for the line source and point source excitations are shown in Figures 6 (c) and (d), respectively, which clearly demonstrate wave focusing on the lens edge (Figures 6 (c)) and wave collimation (Figures 6 (d)). These results are in good agreement with the numerical simulations shown in Figure 3 at a frequency of $f = 40$ kHz.

To further analyse the focusing performance of the lens, the out-of-plane displacements at the focal point along the dashed lines shown in Figure 6 (a) and 6(c) were extracted and compared in Figure 6 (e). The results exhibit a significantly higher (~3 times) displacement amplitude at $y = 0$ for the plate with lens compared to that of the reference plate. Furthermore, this large displacement amplitude was focused only in a narrow range of $-0.01\text{m} < y < 0.01$ m, falling away quickly as $y$ goes beyond this range. These results clearly indicate that the flexural waves were focused onto a small spatial region by the Luneburg lens, as designed.

In addition, the phase data were extracted from Figure 6 (b) and (d) (along the dashed lines, 0.1 m away from the lens center) to study the collimation performance. It can be seen that the phase obtained on the reference plate fluctuates from $-\pi$ to $\pi$, while the phase obtained on the plate with lens remains constant from $y = -0.04$ m to y $= 0.04$ m along y axis, confirming the collimation property of the Luneburg lens. In addition, another modified Luneburg lens ($F/R = 1.5$) was also fabricated and characterized. The measured results in frequency domain are shown in Figure S6, which are in good agreement with the numerical simulations (Figure 3).

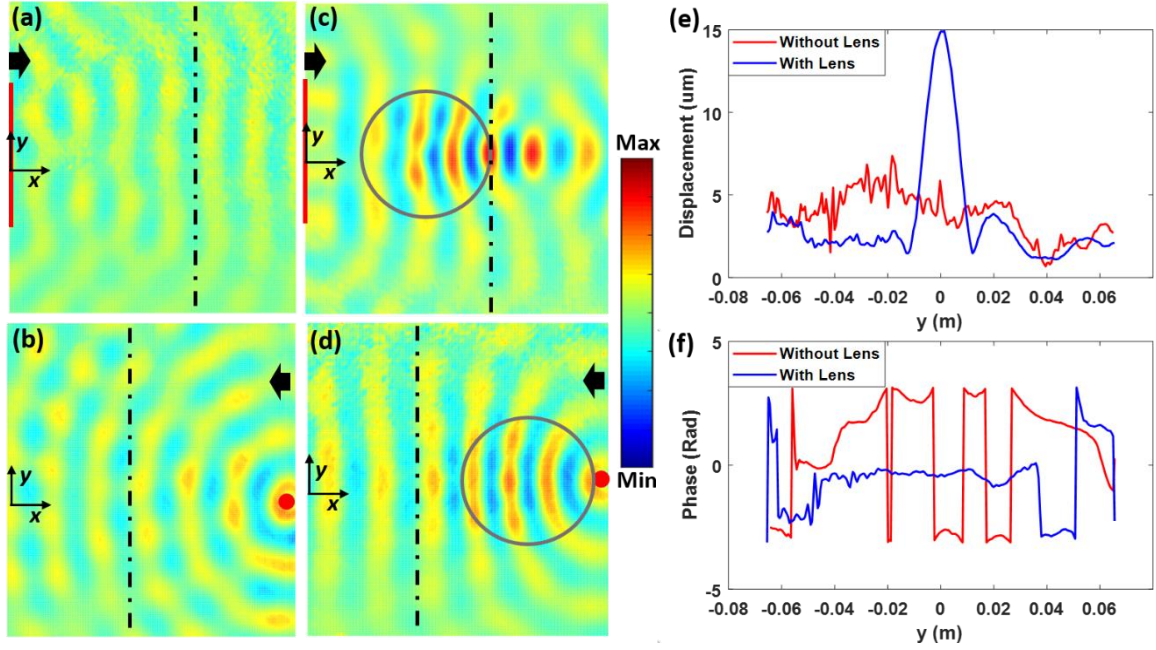

**Figure 6: Experimentally measured steady state response for both focusing and collimation. (a) and (b) are the flexural wave propagation with line source (red line on the left edge of (a)) and point source (red spot on the right edge of (b)) excitations on a reference plate without lens, respectively. (c) and (d) are the corresponding flexural wave focusing (with line source excitation) and collimation (with point source excitation) on the plate with Luneburg lens (focal length ratio of $F/R = 1$). (e) is the displacement data along the dash-dotted lines in (a) and (c) to show the focusing characteristic. (f) is the phase data along the dash-dotted lines in (b) and (d) to show the collimation characteristic. The grey circle indicates the outline of the Luneburg lens, and the color bar indicates the displacement. The black arrows indicate wave propagation directions.**

The transient responses of the flexural wave focusing and collimation were also obtained in experiments for the case of $F/R = 1$, as shown in Figure 7. These results are in excellent agreement with the simulation results shown in Figure 4. For comparison, the time domain measurements of the reference plate are provided in Figure S7. Furthermore, the time domain measurement results were also obtained with another modified Luneburg lens with $F/R$

= 1.5, as shown in Figure S8 of the supplementary material. The results are, once again, consistent with the numerical simulations (Figure S5). These results, taken together, experimentally validates the design flexibility that Equation (4) provides by relating the thickness profile with the focal length ratio.

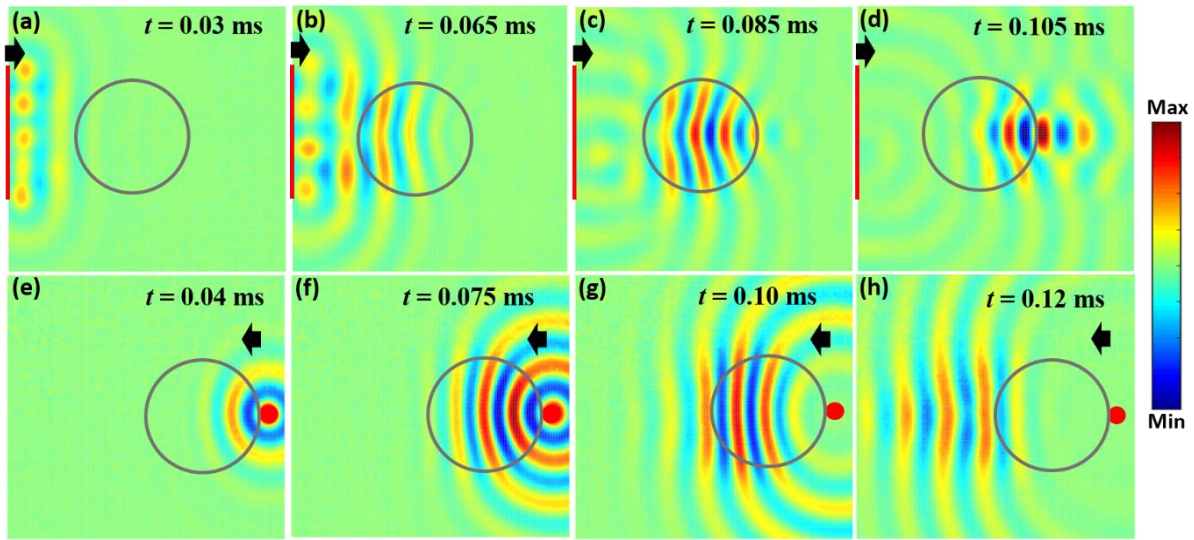

**Figure 7: Measured transient response for both focusing and collimation with the modified Luneburg Lens ($F/R = 1$). (a)-(d) are the flexural wave focusing at time instants of $t$ = 0.03, 0.065 ms, 0.085, and 0.105 ms. The red line indicates the line source excitation. (e)-(h) are the flexural wave collimation at time instants of $t$ = 0.04, 0.075, 0.10, and $t$ = 0.12 ms. The red spot indicates the point source excitation. The grey circle indicates the outline of the Luneburg lens, and the color bar indicates the displacement. The black arrow indicates the wave propagation direction. .**

## 4. Energy Harvesting

As demonstrated in the previous sections, the Luneburg lens can be used to achieve the focusing of flexural waves. The wave energy density at the focal point can be greatly enhanced, which can help improve the performance of vibration-based energy harvesting. In this section, the energy harvesting performance based on a Luneburg lens was investigated numerically and experimentally.

**4.1 Numerical simulations**

The numerical model used here is the same as that described in Section 3.2 for flexural wave focusing, except that a piezoelectric (PZT-5H) energy harvester was attached to the plate (Figure 8 (a)). The material properties of PZT-5H with polarization in the $d_{31}$ direction were used to obtain out-of-plane vibrational energy. Fully coupled electro-mechanical modelling was conducted to simulate the coupling between electrical response and flexural vibrations. In the time domain analysis, Low Reflecting Boundaries (LRB) were used to reduce the boundary reflections. For flexural wave focusing, a line source located at $l = -0.08$ m from the lens centre was used for excitation. The time domain input signal was a 3-count tone bursts at the frequency $f = 40$ kHz. Here, simulations were only performed for the focal length ratio of $F/R = 1$ as similar results are expected for other focal length ratios.

The time domain and frequency domain simulation results are presented in Figure 8 (b) and (c), respectively. In the time domain, the normalized voltage is defined as $\{V_{N1}, V_{N2}\} = \{V_1, V_2\}/\max(V_1)$, where $V_{N1}$ is the normalized voltage obtained with the plate with Luneburg lens and $V_{N2}$ is the normalized voltage obtained with the reference plate, $V_1$ and $V_2$ are the voltages obtained with the piezoelectric energy harvesters on the plates with and without Luneburg lens, respectively, and **max** denotes maximum value. The gain spectrum shown in Figure 8 (c) was obtained by using Fourier transform of the instantaneous voltage obtained in Figure 8 (b), which is defined as $G = \mathcal{F}(V_1)/\mathcal{F}(V_2)$, where $\mathcal{F}$ represents Fourier transform. These numerical results suggest that the Luneburg lens can help achieve enhanced energy harvesting in both time domain and frequency domain over a broadband frequency range (20 – 80 kHz). In particular, the peak energy value in the time domain was increased up to 200% compared with that without the Luneburg lens, and the gain at specific frequency of 58 kHz was ~20.

**4.2 Experimental verifications**

Figure 8 (d) shows the structural Luneburg lens with the line source and the piezoelectric energy harvester. Five rectangular piezoelectric transducers were used collectively as the line source, and a circular piezoelectric disc was used as an energy harvester. As a proof-of-concept, the experiment was conducted on the modified structural Luneburg lens with $F/R = 1$. A uniform plate without Luneburg lens, instrumented with the same types of line source and energy harvester, was used as the reference. The output voltage from the harvester was obtained by using a DAQ system. The experimental results are presented in Figure 8 (e) and (f), which demonstrate enhanced energy harvesting with the Luneburg lens in both the time and frequency domains. The Luneburg lens was able to increase the maximum harvested voltage by up to 300%, and the maximum gain was ~ 24 at 50 kHz. These experimental results are generally consistent with the numerical results, which confirm that the use of the Luneburg lens can enhance the performance of vibration-based energy harvesters.

The slight discrepancy between numerical and experimental results is believed to be mainly due to the material property difference between numerical simulations and experimental setup. In addition, the inevitable inconsistency in the bonding layer of piezoelectric transducers in the experiments may also affect the experimental results. Furthermore, in the numerical simulations, the bonding layer between piezoelectric transducers and aluminium plate was not taken into consideration. Nevertheless, the enhanced energy harvesting performance obtained with the Luneburg lens was found to be similar to other devices, such as Acoustic Black Holes [16, 18] and mechanical metamaterials [26]. However, the Luneburg lens design provides much improved feasibility and flexibility.

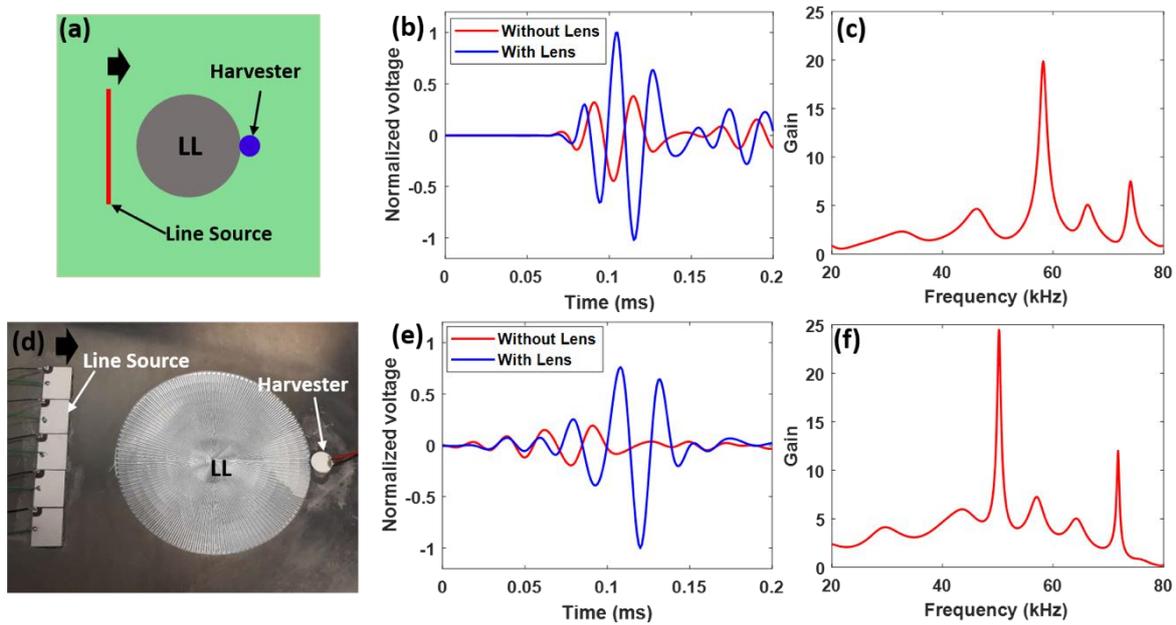

**Figure 8: Numerical and experimental results obtained with Luneburg focusing lens for energy harvesting. (a) Schematic of numerical simulation model. Numerical simulations of normalized voltage in transient response (b) and gain spectrum (c). (d) Experimental setup. (e) Experimental results of normalized voltage in transient response. (f) Experimental results of gain spectrum. Note that the peaks in the gain spectra obtained in (c) and (d) likely correspond to the local resonant modes of the plate with Luneburg lens. The grey area in (a) indicates the Luneburg lens, and the right direction arrow indicates a plane wave excitation direction.**

## 5. Conclusions

In this paper, we proposed a novel design of a modified structural Luneburg lens, which is achieved by using a variable thickness structure. This simple lens design enables the realization of a continuous gradient of refractive index, which allows for the manipulation of flexural wave propagation and overcomes the limitations in existing wave manipulation devices (e.g., singularity in ABHs and discrete lattices in phononic crystals or metamaterial Luneburg lenses). Furthermore, we analytically, numerically, and experimentally demonstrated omnidirectional and broadband flexural wave focusing and collimation with the modified Luneburg lens. Unlike the conventional Luneburg lens, the modified lens allows focusing of flexural wave at different locations via tuning of the thickness profile of the lens, which provides great flexibility in practical applications of the lens. Ray tracing method was used to provide analytical calculations to guide the design of the lens for focusing and collimation, which were validated through numerical simulations and experimental measurements. In addition, the focusing property of the Luneburg lens was shown to help enhance the performance of vibration-based energy harvesting, which was confirmed through both numerical and experimental results. The insights gained through this study are expected to provide new solutions for the development of high-performance acoustic transducers, acoustic waveguides, and vibration-based energy harvesting devices.

**Conflict of Interest**

The authors declare no conflict of interest.